\documentclass[acmtog,nonacm]{acmart}

\usepackage{booktabs} %
\usepackage{multirow}
\usepackage{subcaption}
\usepackage{wrapfig}
\graphicspath{{./}}

\usepackage[normalem]{ulem}

\usepackage[ruled]{algorithm2e} %

\SetAlFnt{\small}
\SetAlCapFnt{\small}
\SetAlCapNameFnt{\small}
\SetAlCapHSkip{0pt}

\acmJournal{TOG}

\begin{document}
\title{\textit{JUST-DUB-IT}:  Video Dubbing via Joint Audio-Visual Diffusion}

\author{Anthony Chen}
\authornote{Equal contribution.}
\authornote{Work done during visit at Tel Aviv University and internship at Lightricks.}
\affiliation{%
  \institution{Tel Aviv University}
  \city{Tel Aviv}
  \country{Israel}
}
\affiliation{%
  \institution{Lightricks}
  \city{Jerusalem}
  \country{Israel}
}
\email{antonchen@outlook.com}

\author{Naomi Ken Korem}
\authornotemark[1]
\affiliation{%
  \institution{Lightricks}
  \city{Jerusalem}
  \country{Israel}
}

\author{Gal Zeevi}
\affiliation{%
  \institution{Lightricks}
  \city{Jerusalem}
  \country{Israel}
}

\author{Tavi Halperin}
\affiliation{%
  \institution{Lightricks}
  \city{Jerusalem}
  \country{Israel}
}

\author{Matan Ben Yosef}
\affiliation{%
  \institution{Lightricks}
  \city{Jerusalem}
  \country{Israel}
}

\author{Urska Jelercic}
\affiliation{%
  \institution{Lightricks}
  \city{Jerusalem}
  \country{Israel}
}

\author{Ofir Bibi}
\affiliation{%
  \institution{Lightricks}
  \city{Jerusalem}
  \country{Israel}
}

\author{Or Patashnik}
\affiliation{%
  \institution{Tel Aviv University}
  \city{Tel Aviv}
  \country{Israel}
}

\author{Daniel Cohen-Or}
\affiliation{%
  \institution{Tel Aviv University}
  \city{Tel Aviv}
  \country{Israel}
}

\makeatletter
\def\@authorsaddresses{}
\makeatother

\begin{abstract}
Audio-Visual Foundation Models, which are pretrained to jointly generate sound and visual content, have recently shown an unprecedented ability to model multi-modal generation and editing, opening new opportunities for downstream tasks.
Among these tasks, video dubbing could greatly benefit from such priors, yet most existing solutions still rely on complex, task-specific pipelines that struggle in real-world settings.
In this work, we introduce a single-model approach that adapts a foundational audio-video diffusion model for video-to-video dubbing via a lightweight LoRA.
The LoRA enables the model to condition on an input audio-video while jointly generating translated audio and synchronized facial motion.
To train this LoRA, we leverage the generative model itself to synthesize paired multilingual videos of the same speaker.
Specifically, we generate multilingual videos with language switches within a single clip, and then inpaint the face and audio in each half to match the language of the other half.
By leveraging the rich generative prior of the audio-visual model, our approach preserves speaker identity and lip synchronization while remaining robust to complex motion and real-world dynamics.
We demonstrate that our approach produces high-quality dubbed videos with improved visual fidelity, lip synchronization, and robustness compared to existing dubbing pipelines.
Project webpage available at \href{https://justdubit.github.io}{\textcolor{red}{https://justdubit.github.io}}.
\end{abstract}

\maketitle

\section{Introduction}

\begin{figure}[t]
    \centering
    \includegraphics[width=\linewidth]{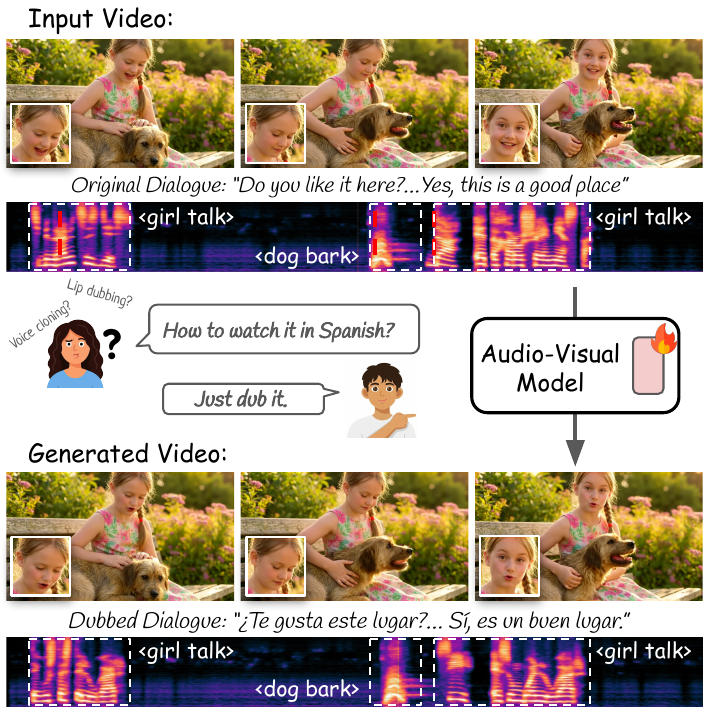} 
    \caption{\textbf{Video dubbing via joint audio–visual generation.}
    \textit{Top}: an input video with spoken dialogue in the source language.
    \textit{Bottom}: the same video dubbed into a target language, generated by our trained model built on top of an audio–visual foundation backbone. Translated speech and lip motion are produced jointly, while the visual context (such as scene dynamics, face expressions, and body movements), speaker identity, and non-speech events (e.g., pauses, background sounds) are preserved.}
    \label{fig:teaser}
\end{figure}

Recent years have seen significant progress in generative models for images~\cite{flux2024, esser2024scalingrectifiedflowtransformers}, video~\cite{wan2025, HaCohen2024LTXVideo}, and audio~\cite{liu2023audioldm, audioldm2-2024taslp, qin2023openvoice}, enabling high-quality content synthesis and editing. These advances have transformed how visual and audio content is created. In particular, it has become increasingly easy to generate and edit media using intuitive, high-level controls such as natural language. More recently, audio-video generative models have emerged, allowing sound and visual content to be modeled jointly in a single generative process~\cite{ltx2, low2025ovi}.

Video dubbing aims to translate the spoken content of a video into a different language while preserving the speaker's identity, including both facial appearance and voice, and maintaining accurate lip synchronization. Unlike audio-only translation or voice conversion, dubbing requires modifying facial motion in a highly localized and precise manner while keeping the rest of the video visually consistent. This makes the task particularly challenging, as even small errors in lip motion, voice characteristics, or identity cues are easily perceived, especially in real-world videos with complex motion, pose changes, and varying visual conditions.

Most existing video dubbing methods decompose the problem into a sequence of specialized stages, typically combining audio synthesis with audio-driven facial editing~\cite{li2024latentsync, zhang2024musetalk, yang2025infinitetalk}. While effective in controlled settings, these modular pipelines are often complex and brittle, with each component making strong assumptions about the input. In practice, such assumptions can break down in unconstrained videos, for example when the speaker moves relative to the camera, produces non-speech sounds such as laughter or sighs, or partially occludes the mouth during speech. Furthermore, most modular systems attempt to isolate the speaker’s voice from the background audio to perform translation in a vacuum. This separation-and-remixing strategy fails to account for the temporal and semantic dependencies between the speaker and their environment. If the translated speech differs in duration from the original—a common occurrence in dubbing—simply re-layering the original background audio leads to a loss of synchrony with environmental events, such as a dog barking in response to a speaker or a door slamming at a specific moment in a conversation. These ``messy'' auditory-visual misalignments break the viewer’s immersion and highlight the limitations of treating speech as a detached signal. In short, separating audio and visual editing prevents the two modalities from informing each other, which limits performance in complex, real-world videos. 

To address these challenges, we take a different approach and frame video dubbing as a joint audio-video generation problem within a single model. Instead of relying on multiple specialized components, we adapt a foundational audio-visual diffusion model to perform dubbing directly, allowing audio and visual cues to be generated together and to inform each other in cross-modality attention layers. 
Crucially, this holistic modeling ensures that the interaction between the speaker and the rest of the scene is preserved. By treating the entire audio-visual stream as a single generative task, our model can naturally adjust the timing and placement of environmental sounds to remain coherent with the newly generated speech and facial motion. This joint formulation naturally captures correlations between speech, facial motion, and scene dynamics, while avoiding many of the assumptions and failure modes of modular pipelines. Importantly, the adaptation is lightweight, requiring only a small LoRA~\cite{hu2021loralowrankadaptationlarge} fine-tuning on top of a strong generative prior, which makes the approach simple, flexible, and robust in practice.

A key challenge in adapting a generative audio-video model for dubbing is the lack of paired training data that preserves both the speaker's identity and the original visual content while allowing the spoken language to change. To address this, we generate training data using the generative model itself. We first generate multilingual videos in which the same speaker switches languages within a single clip, ensuring consistent facial appearance and voice characteristics across languages. We then split the clip into two halves and each time inpaint the face and audio in one half to match the language of the other half, while keeping the underlying semantic content intact. Using audio-video inpainting framework allows the model to leverage both visual and audio context, resulting in aligned bilingual video pairs that provide effective supervision for training.

We evaluate our approach on a diverse set of videos and language pairs, including unconstrained real-world scenarios with complex motion and visual variability. Our method consistently produces high-quality dubbed videos that preserve both facial and voice identity while maintaining accurate lip synchronization and the temporal-semantic coherence of interactions within the scene. Compared to existing dubbing pipelines, our approach is more robust to challenging conditions such as non-frontal views, partial occlusions, and expressive facial behaviors, resulting in improved visual fidelity and overall perceptual quality. These results demonstrate the benefit of leveraging a strong joint audio-video generative prior for dubbing.

\section{Related Work}
\label{sec:related_work}

\paragraph{\textbf{Audio-Visual Generative Models}}
The field of multimodal synthesis is shifting from cascaded pipelines to unified Audio-Visual foundation models, using Diffusion Transformers (DiTs)~\cite{low2025ovi, zhang2025uniavgen, Peebles2022DiT, ltx2}. High-fidelity generation and modality alignment are achieved through architectures like \textsc{Ovi}~\cite{low2025ovi} and \textsc{UniAVGen}~\cite{zhang2025uniavgen}, which employs twin-backbone designs and Asymmetric Cross-Modal Interaction (ATI). Further frameworks, including \textsc{Seedance 1.5 pro}, \textsc{MM-Sonate}, and \textsc{Syncphony}, enhance alignment via joint denoising, flow-matching, and specialized motion-aware normalization~\cite{seedance2025, qiang2026sonate, song2025syncphony, cheng2025mmaudio}. Our work builds upon these priors for end-to-end audiovisual dubbing.

\paragraph{\textbf{Audio-driven Talking Face Generation}}
Audio-driven talking face synthesis evolved from early graphics-based concatenation~\cite{bregler1997video} to deep neural approaches like Wav2Lip, which introduced synchronization discriminators but suffered from texture artifacts~\cite{prajwal2020wav2lip, guan2023stylesync}. Modern approaches can be classified as inpainting-based (e.g., MuseTalk~\cite{zhang2024musetalk}, LatentSync~\cite{li2024latentsync}), which reconstruct masked lip regions but often produce artifacts and discontinuities~\cite{peng2025omnisync, yaman2024audiodriventalkingfacegeneration}, and non-inpainting methods like InfiniteTalk, OmniSync, and X-Dub, which avoid explicit masking and directly edit videos using Video DiTs~\cite{yang2025infinitetalk, peng2025omnisync, he2025xdub}. Our method uniquely synthesizes paired cross-lingual dubbing datasets, training a unified model that comprehensively preserves both audio and visual contexts of the original content.

\paragraph{\textbf{Zero-shot Voice Cloning}}
Voice cloning, a crucial component in traditional dubbing pipelines, has transitioned from data-intensive methods to zero-shot models leveraging neural codec language models and large language model adaptations~\cite{wang2025Neural, ye2025llasascalingtraintimeinferencetime, cui2025glmtts, ju2024natural}. Approaches like IndexTTS2 and SSPO specifically ensure precise duration alignment~\cite{zhou2025indextts2breakthroughemotionallyexpressive, Cui_2025_sspo}, yet remain unimodal, neglecting vital visual cues essential for accurately rendering paralinguistic elements (e.g., laughter, sighs, breathing). In contrast, our audiovisual generation method leverages visual dynamics to faithfully reproduce these non-verbal expressions and preserve the speaker's original vocal identity.

\section{Method}
Our method adapts a pretrained audio–visual diffusion model for video dubbing without relying on masks, explicit face tracking, or modular pipelines. Instead of using the model directly for dubbing, we leverage its generative capacity to synthesize identity-consistent bilingual training pairs, and then learn a constrained editing behavior on top of it. This section describes the underlying audio–visual foundation model (Sec.~\ref{sec:preliminary}), the construction of paired dubbing data (Sec.~\ref{sec:data}), and the lightweight in-context LoRA that enables precise, temporally aligned dubbing at inference time (Sec.~\ref{sec:train}).

\subsection{Preliminary}
\label{sec:preliminary}

Our approach is built upon LTX-2~\cite{ltx2}, a foundation model that processes video and audio as a unified signal. LTX-2 employs an Asymmetric Dual-Stream Diffusion Transformer (DiT) that processes decoupled latent inputs: video frames are compressed into 3D spatiotemporal tokens $z_v$ via a 3D VAE, and audio is encoded into 1D tokens $z_a$ via a separate 1D VAE. To manage the distinct information densities of these modalities, the model allocates different capacities to each stream, enforcing tight temporal alignment through bidirectional cross-attention layers that allow each modality to continuously condition the other~\cite{ltx2}.

Similar to advanced image diffusion models like Flux~\cite{flux2024}, LTX-2 is trained using Flow Matching (specifically Rectified Flow) ~\cite{lipman2023flowmatchinggenerativemodeling, liu2022flowstraightfastlearning}, which learns straight trajectories between the data and noise distributions. The training process defines a linear probability path $x_t = (1-t)x_0 + t x_1$ connecting the Gaussian noise distribution $x_0 \sim \mathcal{N}(0, I)$ to the data distribution $x_1 \sim \mathcal{D}$. The model is optimized to predict the velocity field $v_\theta$ that drives this transformation by minimizing the regression loss:
\begin{equation}
    \mathcal{L}_{FM} = \mathbb{E}_{t \sim \mathcal{U}[0,1], x_0, x_1} \left[ \| v_\theta(x_t, t, c) - (x_1 - x_0) \|^2 \right],
\end{equation}

where $c$ represents conditioning signals such as text prompts. In LTX-2, the text prompt is highly detailed and contains the full transcription of the audio; thus, the model is not responsible for the translation itself, as the translated prompt is provided directly as conditioning. We adapt this robust, flow-based audio-visual prior for the video dubbing task using Video In-Context Low-Rank Adaptation (IC-LoRA)~\cite{ltx2_codebase}, allowing to steer the synchronized generation process toward target languages while requiring only a minimal number of trainable parameters and a small amount of task-specific training data.

\subsection{Video-dubbing Paired Data Construction}
\label{sec:data}

\begin{figure}[t]
    \centering
    \includegraphics[width=1.0\linewidth]{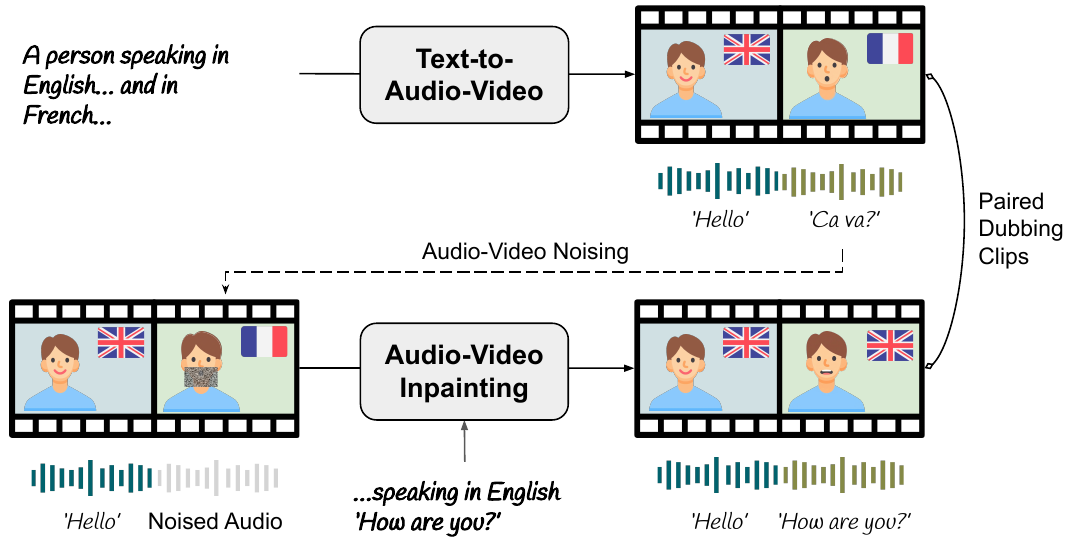}
    \caption{\textbf{Pipeline for Generating Paired Audio-Visual Dubbing Data.}
    The pipeline consists of two stages.  First, the audio–visual generation model produces a contiguous sequence containing a context clip (e.g., spoken English) followed by a target clip (e.g., spoken French).
    Second, the audio and lip-region video of the target clip are masked, and the same unified model is used in an inpainting setting to regenerate the masked content, conditioned on the context clip and a new text prompt (e.g., re-dubbing the target into English).}
    \label{fig:dataset}
\end{figure}

\begin{figure}[t]
    \centering
    \includegraphics[width=1.0\linewidth]{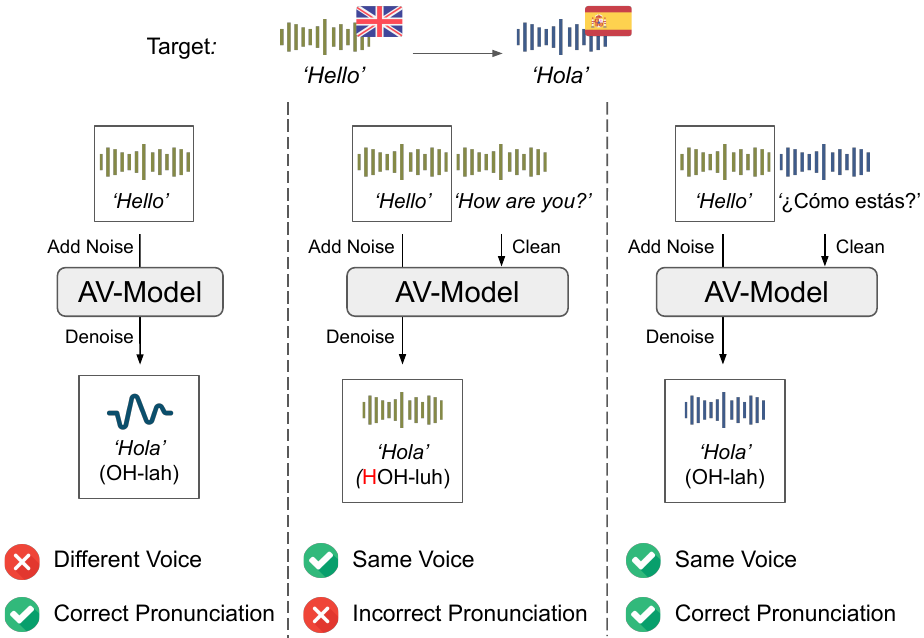}
    \caption{\textbf{The Identity–Pronunciation Trade-off.} Naïve audio inpainting reveals a fundamental conflict between preserving speaker identity and achieving linguistically correct pronunciation. When denoising from scratch (Left), the model exhibits \emph{voice drift}, failing to preserve the speaker’s vocal identity. When conditioning on the source audio to maintain identity (Middle), phonetic and prosodic patterns leak across languages, resulting in \emph{prosody leakage}. Our approach (Right) resolves this trade-off by conditioning generation on a reference clip that preserves speaker identity while exhibiting the target-language phonetic style.}
    \label{fig:audio_tradeoff}
\end{figure}

The core challenge in training a joint audio-visual dubbing model is the lack of ``perfect pairs'': datasets where the same speaker utters the same semantic content in multiple languages while maintaining identical pose, lighting, identity and backgrounds. Since natural data of this nature does not exist, we synthesize it by leveraging the generative prior of a pretrained audio-visual foundation model. {Importantly, although our dataset largely consists of pairs of identical content in different languages, we find that it is effective for training the model to perform a wide range of editing tasks. By learning to map diverse text prompts to synchronized audio-visual updates, the model generalizes at inference time to diverse dubbing tasks, including same-language content editing and translation between previously unseen language pairs.

As illustrated in Fig.~\ref{fig:dataset}, our pipeline operates in two stages. First, we generate Language-Switching Videos—clips where a single speaker naturally transitions between languages (e.g., English $\rightarrow$ French) within a contiguous shot. This establishes a ground-truth reference for the speaker's identity in both linguistic contexts. Second, we employ inpainting to create counterfactual pairs. We split the video at the language switch; for the first half (Language A), we noise the relevant face area and audio, then denoise them conditioned on the translated text and the global audiovisual context of the second half (Language B), including visual identity, voice characteristics, and environmental cues. This yields a dataset of context-aware paired dubbing data, where the visual context (pose, background) remains intact while the spoken language varies. Below we discuss the motivation of our design and three key mechanism we use to improve data quality.

\paragraph{\textbf{The Identity–Pronunciation Trade-off.}}
The language-switching strategy balances vocal identity and pronunciation during audio inpainting. As shown in Fig.~\ref{fig:audio_tradeoff}, denoising without context leads to \textit{voice drift}, while conditioning solely on source-language audio preserves identity but causes \textit{Prosody Leakage} where the target text is articulated with the source language's rhythm. Our approach resolves this by conditioning generation on a reference clip that provides the speaker's vocal identity within the correct target linguistic style.

\paragraph{\textbf{Latent-Aware Fine Masking}}
A critical challenge in training-free inpainting is preventing the model from ``peeking'' at the original motion. A naïve approach involves encoding the original video and then applying noise to the latent tokens corresponding to a downsampled lip mask. However, this strategy fails due to the large receptive field of the video VAE encoder (e.g., $32 \times 32 \times 8$). During the encoding process, spatial information is not strictly isolated; pixel data from the lip region propagates into the latent representations of the surrounding face (e.g., jaw and cheeks). Consequently, even when the specific lip tokens are fully corrupted with noise, the adjacent unmasked tokens retain ``echoes'' of the original lip trajectory. The diffusion model exploits this leakage to trivially reconstruct the original motion rather than generating diverse, audio-aligned dubbing (see visualization in Fig.~\ref{fig:lip_augmentation}). To eliminate this, we implement Latent-Aware Fine Masking. We empirically determine the ``Effective Latent Mask'' by computing the residual difference in latent space between a masked and unmasked input. This identifies and masks the full extent of information spread, forcing the model to regenerate the relevant facial regions from scratch in sync with the generated audio.

\paragraph{\textbf{Lip Augmentation via Phonetic Diversity}}
A secondary challenge in synthetic data generation is the ambiguity of visemes, where different phonemes may share similar lip shapes, often resulting in visually indistinct or “mumbling” lip movements (see Fig.~\ref{fig:latent_masking}). To increase visual discrimination between the condition and the target, we augment the training data by prompting exaggerated, character-level articulation (e.g., “A..B..C”), encouraging more diverse lip movements. In practice, we perform video inpainting twice for each clip: once without Lip Augmentation to obtain correct translated audio, and once with Lip Augmentation to generate visually diverse lip movements. The two outputs are merged into a single video and used as the training context. Although the visual and audio streams are not synchronized at this stage, the impact of such mismatch is resolved using Modality-Isolated Cross-Attention, as described in the following section.

\paragraph{\textbf{Quality Control Pipeline}}
To ensure training fidelity, we employ a multi-stage filtering process (detailed explanations provided in Appendix Sec.~\ref{sec:appendix_data}).

\subsection{Audio-Video In-Context Learning}

\begin{figure}[t]
    \centering
    \includegraphics[width=0.9\linewidth]{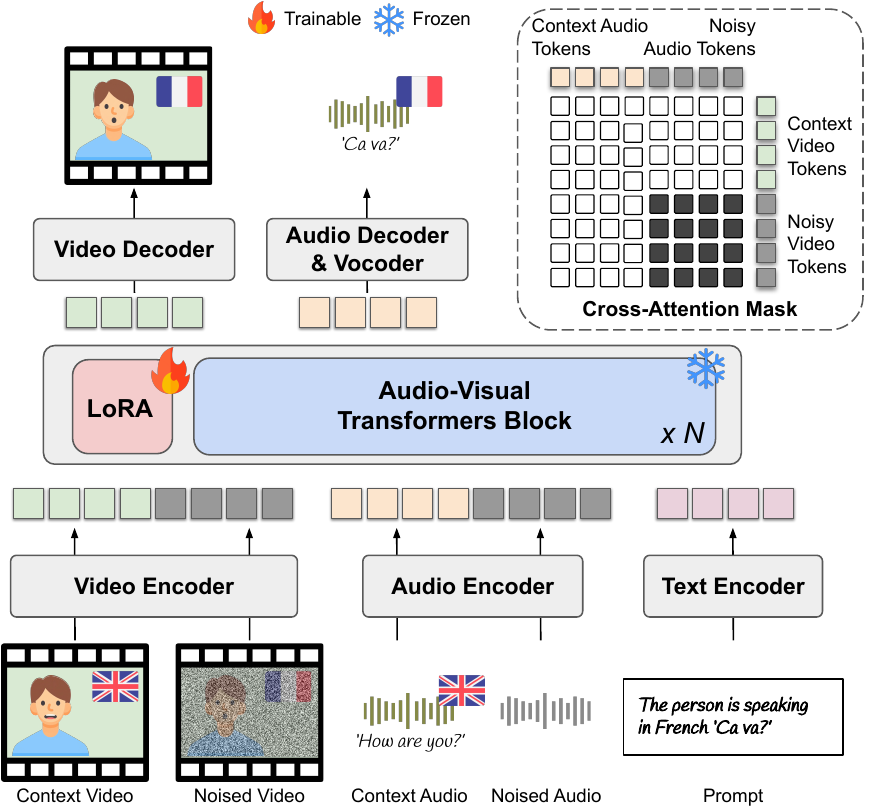}
    \caption{\textbf{Model Training.} Our framework follows an in-context generation paradigm where clean context audio-visual pairs are concatenated with noised target pairs. We fine-tune only LoRA adapters while keeping a pre-trained Audio-Visual (AV) Diffusion Transformer frozen. Conditioned on a text prompt (e.g. ``The person is speaking in French''), the model learns to propagate edits from the context while maintaining temporal synchronization between audio and video. We introduce a modality-specific masking strategy in AV cross-attention, ensuring that noisy audio attends only to noisy video and vice versa, since conditioning a noisy signal on clean context from the other modality leads to signal leakage and conflicting guidance, which this masking prevents.}
    \label{fig:model}
\end{figure}
    
\paragraph{\textbf{Audiovisual In-Context Framework}}
\label{sec:train}

Inspired by the In-Context LoRA (IC-LoRA) paradigm~\cite{lhhuang2024iclora}, we introduce a framework for joint video audio editing that enables temporally synchronized and context aware generation. We formulate the editing process as a conditional generation task: given a source audiovisual pair $(V_{src}, A_{src})$ and a desired text dialogue, the model learns to edit the visual and audio content(i.e. lips and the related speech) to obtain a lip-synchronized audiovisual pair that matches the target dialogue, while preserving the context from the source, including visual and voice identity, as well as the environment.

To achieve this without the prohibitive cost of full parameter fine-tuning, we adopt a Low-Rank Adaptation (LoRA) approach. We inject trainable rank-decomposition matrices into the self-attention and cross-attention layers of a pre-trained audiovisual transformer. During training, the base model weights remain frozen, and only the LoRA adapters are optimized to adapt the velocity field prediction by minimizing the flow-matching objective:
\begin{equation}
\mathcal{L}_{FM} = \mathbb{E}_{t, x_0, x_1, c} \left[ \| v_\theta(x_t, t, c, \text{LoRA}) - (x_1 - x_0) \|^2 \right].
\end{equation}
where $c$ represents the source visual tokens, audio tokens and text tokens.

\paragraph{\textbf{Context-Aligned Multimodal Positional Encoding.}}
Shared positional encodings are important for maintaining intra-modal alignment. For both video and audio streams, we assign contextual tokens positional encodings identical to their target counterparts. This explicitly signals to the model that contextual frames and audio segments are spatially and temporally aligned with the target, reframing dubbing as a context-aware completion task grounded in the original scene.

By sharing positional encodings across contextual and target modalities, we guide the LoRA layers to learn mappings that preserve the spatial and temporal structural integrity of the scene while steering generation toward the target language and speaker identity. This design reframes dubbing as a context-aware completion task rather than a blind translation process, enabling the model to generate lips and speech that remains grounded in the original scene and maintains fine-grained audiovisual synchronization.

\paragraph{\textbf{Modality-Isolated Cross-Attention}}
A notable hurdle in joint audiovisual modeling involves managing the dense interactions within transformer blocks. While cross-attention layers are essential for aligning video and audio features, a global attention mechanism in an in-context setting often results in \textit{cross-modal leakage}. Specifically, unnecessary interactions between the \textbf{source} video tokens and the \textbf{noisy} target audio tokens (and vice versa) can introduce guidance noise, leading to blurred boundaries or temporal misalignments in the final output.

To address this leakage issue, we introduce a structured attention bias within cross-modality attention layers. For a query $Q$ from one modality and keys $K$ from both source and target contexts, the attention operation is defined as:
\begin{equation}
\text{Attention}(Q, K, V) = \text{Softmax}\left(\frac{QK^T}{\sqrt{d_k}} + M\right)V,
\end{equation}

where $M$ is a masking matrix designed to restrict the interaction space. Specifically, we define $M$ such that:
\begin{equation*}
M_{i,j} =
\begin{cases}
0 & \text{if } i,j \in \text{Target (noisy) tokens of both modalities} \\
-\infty & \text{otherwise}
\end{cases}
\end{equation*}

This masking strategy forces the denoising process to derive cross-modal alignment solely from the target pair while permitting each modality to reference its respective source tokens for identity guidance. This approach resolves the audiovisual asynchrony introduced by lip augmentation, ensuring each modality focuses exclusively on relevant cross-modal signals.

Our unified model enables visual dynamics and acoustic events to co-evolve naturally. By treating the entire stream as a single generative task, the model synchronizes paralinguistic cues—such as sighs or laughter—and grounds environmental sounds (e.g., a dog barking) in the physical actions of the scene. This holistic approach ensures dialogue pacing remains grounded in scene context, avoiding the temporal misalignment common in modular pipelines that treat speech as a detached stream.

\section{Experiments}
\label{sec:experiments}

\begin{table}[t]
\centering
\caption{\textbf{Quantitative Evaluation of Visual Quality and Audiovisual Synchronization.} 
We report generation success rate (Succ), identity preservation (CSIM), visual fidelity (FID), temporal coherence (FVD), Mouth Aspect Ratio diversity (MAR Div.), and audiovisual synchronization (ASync).}
\label{tab:visual_sync_qualty}
\vspace{-2mm}
\resizebox{\linewidth}{!}{
\begin{tabular}{l | c cccc c c}
\toprule
\textbf{Method} 
& Succ $\uparrow$ 
& CSIM $\uparrow$ 
& FID $\downarrow$ 
& FVD $\downarrow$ 
& MAR Div. $\uparrow$
& ASync $\downarrow$ \\
\midrule

\multicolumn{7}{c}{\textbf{HDFT \& TalkVid Benchmark}} \\
\midrule
Real Data & -- & -- & -- & -- & 0.1263 & 1.2900 \\
\midrule
LatentSync + CosyVoice 
& \textbf{100\%} 
& \textbf{0.8762} 
& \textbf{2.63} 
& \underline{260.98} 
& \underline{0.1043} 
& \textbf{0.05} \\
MuseTalk + CosyVoice 
& \textbf{100\%} 
& 0.7937 
& \underline{4.66} 
& 272.96 
& 0.0988 
& \underline{0.70} \\    
\midrule
\textbf{Ours (Unified)} 
& \textbf{100\%} 
& \underline{0.8471} 
& 8.45 
& \textbf{131.88} 
& \textbf{0.1168} 
& 2.21 \\

\midrule \midrule

\multicolumn{7}{c}{\textbf{Challenging Benchmark}} \\
\midrule
Real Data & -- & -- & -- & -- & 0.1416 & 2.4433 \\
\midrule
LatentSync + CosyVoice 
& \underline{80\%} 
& \textbf{0.7070} 
& \textbf{6.45} 
& \underline{758.66} 
& \underline{0.1335} 
& \textbf{1.34} \\
MuseTalk + CosyVoice 
& 74\% 
& 0.5770 
& 16.04 
& 902.03 
& 0.1131 
& 5.65 \\
\midrule
\textbf{Ours (Unified)} 
& \textbf{100.0\%} 
& \underline{0.6455} 
& \underline{12.45} 
& \textbf{353.54} 
& \textbf{0.1461} 
& \underline{2.44} \\
\bottomrule
\end{tabular}
}
\end{table}

\begin{table}[t]
\centering
\caption{\textbf{Quantitative Audio Quality Comparisons.} We evaluate temporal alignment (Dur-Err), voice similarity (V-SIM), audio intensity consistency (Int-Corr), and linguistic accuracy (WER) against audio-only baselines.}
\label{tab:audio_quality}
\vspace{-2mm}
\resizebox{\linewidth}{!}{
\begin{tabular}{l l | c c c c}
\toprule
\textbf{Dataset} &
\textbf{Method}
& Dur-Err (s) $\downarrow$
& V-SIM $\uparrow$
& Int-Corr $\uparrow$
& WER $\downarrow$ \\
\midrule
\multirow{3}{*}{\begin{tabular}{l}
HDFT \& \\
TalkVid \\
Benchmark
\end{tabular}}
& CosyVoice
& 3.4995
& \textbf{0.6868}
& 0.416
& 0.3288 \\
& OpenVoice
& \underline{1.1628}
& 0.4006
& \underline{0.506}
& \textbf{0.1725} \\
& \textbf{Ours (Unified)}
& \textbf{0.1638}
& \underline{0.4235}
& \textbf{0.758}
& \underline{0.3133} \\
\midrule
\multirow{3}{*}{\begin{tabular}{c}
Challenging \\
Benchmark
\end{tabular}}
& CosyVoice
& \underline{1.9612}
& \textbf{0.6168}
& 0.478
& \underline{0.1573} \\
& OpenVoice
& 1.9917
& 0.3605
& \underline{0.501}
& \textbf{0.0630} \\
& \textbf{Ours (Unified)}
& \textbf{0.0463}
& \underline{0.5578}
& \textbf{0.815}
& 0.2739 \\
\bottomrule
\end{tabular}
}
\vspace{-4mm}
\label{table_2}
\end{table}

\begin{figure}[t]
    \centering
    \includegraphics[width=\linewidth]{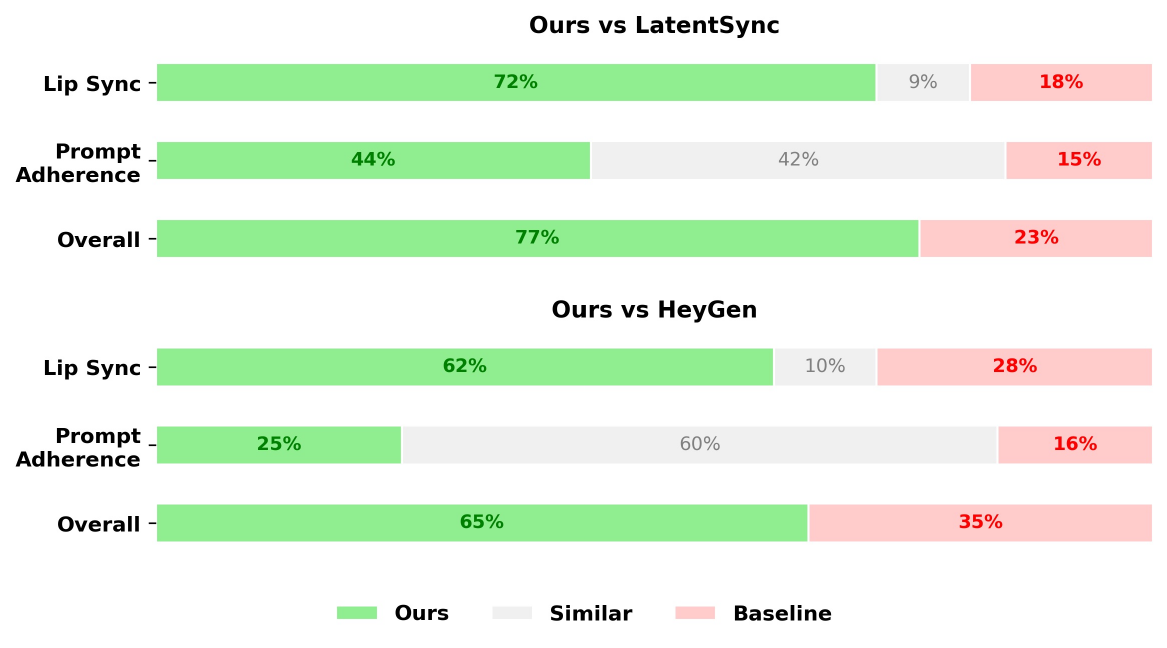}
    \vspace{-6mm}
    \caption{\textbf{User Study Results.} We compare our method against LatentSync and HeyGen through a user study, evaluating Lip Synchronization, Prompt Adherence, and Overall Quality. Results indicate that participants prefer our method over baselines across all evaluated metrics.}
    \label{fig:user_study}
    \vspace{-4mm}
\end{figure}

\begin{figure*}[ht]
    \centering
    \begin{subfigure}[t]{0.49\linewidth}
        \centering
        \includegraphics[width=\linewidth]{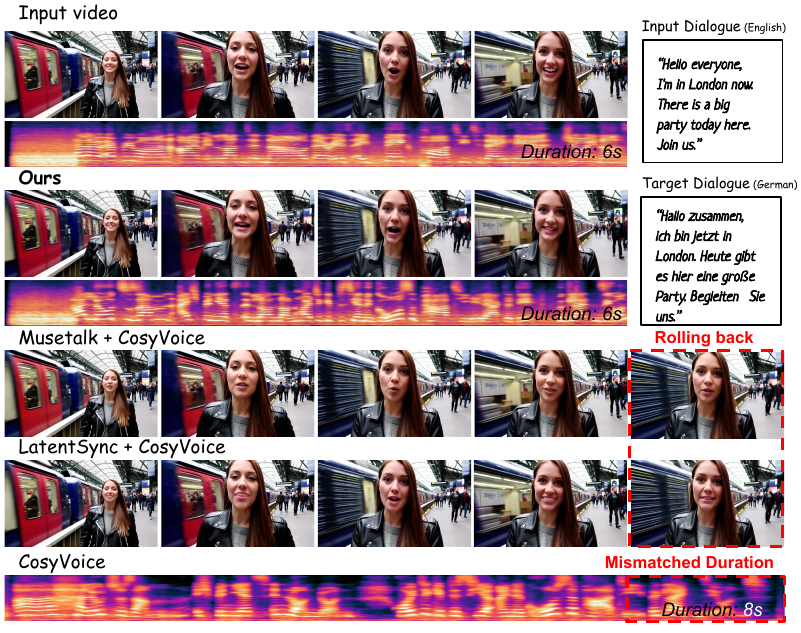}
        
        \caption{\textbf{Duration Alignment.} Our method precisely matches the original video’s duration and synchronizes translated audio, effectively avoiding baseline artifacts such as reversed motion (“Rolling back”) or mismatched durations.}
        \label{fig:comp_duration}
    \end{subfigure}
    \hfill
    \begin{subfigure}[t]{0.49\linewidth}
        \centering
        \includegraphics[width=\linewidth]{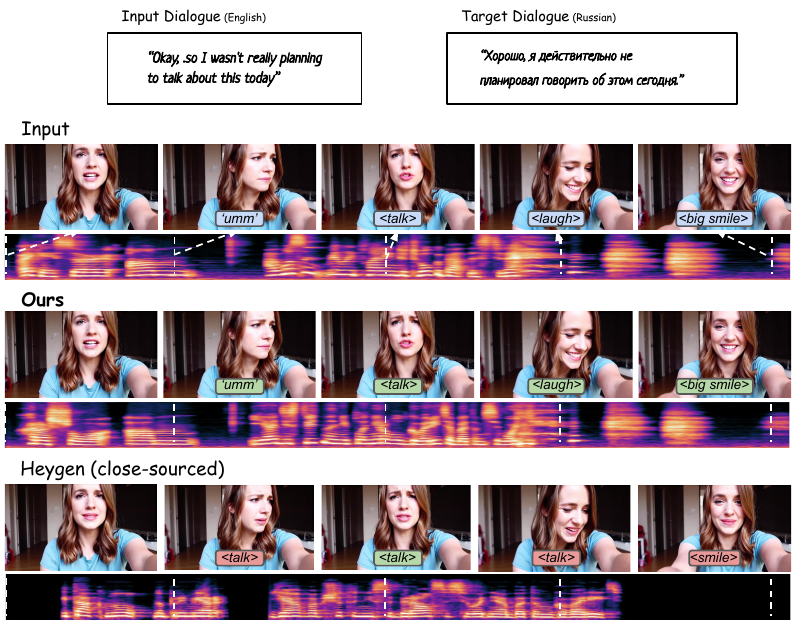}
        
        \caption{\textbf{Preservation of Non-Dialogue Events.} Our method naturally preserves crucial non-dialogue audio events (e.g., pauses, laughter), leveraging visual cues, while baseline methods ignore these segments entirely.}
        \label{fig:comp_nondialogue}
    \end{subfigure}
    \caption{\textbf{Qualitative Comparisons on Audiovisual Alignment and Temporal Structure} Our joint audiovisual model accurately maintains both visual and auditory contexts, outperforming baseline methods on key audiovisual synchronization tasks.}
    \label{fig:comp_combined}
\end{figure*}

\paragraph{\textbf{Datasets}}
We evaluate our framework using a combination of standard and challenging benchmarks. Following previous works, we adopt the widely used HDFT~\cite{zhang2021flow} and TalkVid~\cite{chen2025talkvidlargescalediversifieddataset} datasets, comprising well-curated videos characterized by predominantly frontal faces, limited pose variation, and clean acoustic conditions. These datasets thus represent relatively narrow-domain evaluation settings. From the test partitions of these datasets, we randomly sample 150 videos to construct English-to-Foreign (covering Russian, Spanish, German, and French), Foreign-to-English, and English-to-English pairs. To complement these benchmarks and better reflect real-world deployment scenarios, we further introduce a more challenging evaluation set by collecting 25 real videos from YouTube and 25 synthetic videos, exhibiting profile views, significant pose shifts, occlusions, and stylized appearances, enabling a more rigorous assessment of robustness under unconstrained conditions.

\paragraph{\textbf{Baselines}}Our framework is the first to jointly perform \textit{visual dubbing} and \textit{audio dubbing} within a single, unified audiovisual generative model. As no prior methods exist that simultaneously address both tasks, we separately compare our approach against state-of-the-art methods from each domain. Specifically, for the \textit{visual dubbing} task, we evaluate MuseTalk~\cite{zhang2025musetalkrealtimehighfidelityvideo} and LatentSync~\cite{li2024latentsync}. For cross-lingual visual dubbing, we use CosyVoice~\cite{du2024cosyvoice} to generate identity-preserving translated audio for the video-only baselines. For the \textit{audio dubbing} task, we compare our method directly against leading zero-shot voice cloning methods, specifically CosyVoice and OpenVoice~\cite{du2024cosyvoice, qin2023openvoice}.

\paragraph{\textbf{Evaluation Metrics}}
We concisely evaluate three dimensions, briefly described below (detailed explanations provided in Appendix Sec.~\ref{sec:appendix_metric}):

Video Quality: Generation success rate (Succ) measures the system's ability to produce an output video. A failure is defined as a complete lack of output, typically occurring when mask-based baselines fail to detect a face, whereas any generated video is counted as a success.), Visual fidelity (FID~\cite{fid}), temporal coherence (FVD~\cite{fvd}), and identity preservation (ID-SIM, ArcFace~\cite{deng2019arcface}) and facial articulation measured by Mouth Aspect Ratio diversity (MAR).

Audio Quality: Linguistic accuracy (WER, Whisper~\cite{opeaniwhisper}), voice similarity (V-SIM, ERes2Net~\cite{zhou2021eres2net}), and temporal intensity consistency (Int-Corr, Pearson correlation of RMS audio envelopes~\cite{chung2024t}).

Audiovisual Synchronization: Temporal offset between lip movements and speech (SyncNet~\cite{syncnet}).

\subsection{Quantitative Results}
\label{sec:result_quanti}

We evaluate our unified audiovisual generation framework across video quality, audio quality, and audiovisual synchronization. Quantitative results are reported in Tab.~\ref{tab:visual_sync_qualty}, Tab.~\ref{tab:audio_quality}, and Fig.~\ref{fig:user_study}.

\paragraph{\textbf{Visual Quality Analysis.}} 
As shown in Tab. ~\ref{tab:visual_sync_qualty} our method achieves the lowest FVD across all datasets, indicating superior temporal coherence and reduced motion artifacts, particularly in challenging in-the-wild scenarios. This robustness is reflected in our 100\% generation success rate on both standard and challenging benchmarks, whereas modular baselines like LatentSync and MuseTalk drop to 80\% and 74\%, respectively, on complex samples. As illustrated by the red 'ERROR' frames in Fig.~\ref{fig:comp_quality}, these failures typically occur when mask-based baselines are unable to detect a face in profile views or non-human subjects, resulting in a total lack of output. Regarding identity preservation, our Unified Model remains highly competitive, we achieve a CSIM of 0.8471 on the HDFT benchmark and 0.6455 on the Challenging Benchmark, scores that are comparable to state-of-the-art modular pipelines. However, as traditional metrics like SyncNet may reward distorted, frontal-biased reconstructions in profile views (see explanation in Fig.~\ref{fig:async_overfit}), the qualitative superiority of our method's temporal-semantic coherence is most apparent when assessing the synchronous audiovisual dynamics. Our FID is higher than face-centric baselines, this is an expected consequence of our global VAE-based encoder–decoder, which operates on full frames rather than masked facial regions. Although this increases reconstruction variance and negatively affects FID, it enables robust generalization to arbitrary characters, non-frontal views, and non-human subjects without requiring face masks or landmarks. Additionally, our Unified Model achieves a higher MAR, demonstrating improved facial expressiveness and natural mouth articulation compared to modular baselines such as MuseTalk and LatentSync, which show restricted movements and "mumbling" artifacts typical of inpainting-based methods.

\paragraph{\textbf{Audio Quality Analysis.}}
As shown in Tab. ~\ref{tab:audio_quality}, our Unified model achieves the lowest duration error (Dur-Err) and highest intensity correlation (Int-Corr). Unlike CosyVoice and OpenVoice, which lack the visual grounding to adjust speech speed to the original shot length, our model naturally compresses or expands dialogue to match the video context. This prevents common artifacts like mismatched endings or "rolling back" motion (see Fig.~\ref{fig:comp_duration}). Our joint formulation ensures that paralinguistic cues and scene interactions—such as a dog barking in response to a gesture—co-evolve with the visual stream (see Fig.~\ref{fig:comp_duration} and Fig.~\ref{fig:comp_event2}). While these advantages are reflected in our superior Int-Corr scores, this temporal-semantic coherence is best experienced through the continuous audiovisual playback. While our Word Error Rate (WER) and voice similarity (SSIM) scores remain competitive, they are inherently bounded by the base audio-visual model's capabilities. Overall, these results highlight our method's strong suitability for in-the-wild audiovisual coherence and realistic speech synchronization in diverse scenarios.

\paragraph{\textbf{Audiovisual Synchronization.}}
Our audiovisual synchronization, measured by AV offset using SyncNet (See Tab.~\ref{tab:visual_sync_qualty}), remains consistently within a perceptually unnoticeable range (approximately 1–2 frames). Although LatentSync and MuseTalk achieve near-zero offsets due to explicit optimization with SyncNet on frontal-view samples, this suggests potential overfitting rather than broadly generalizable synchronization performance. In contrast, our method demonstrates synchronization quality highly competitive with real videos, effectively capturing natural lip movements across varied and challenging scenarios. 

\paragraph{\textbf{User Study}}
Our user study (Fig.~\ref{fig:user_study}) confirms that our method is preferred over baselines—including the SOTA closed-source solution from HeyGen~\cite{heygen2026}, in terms of \textit{Lip Synchronization}, 	\textit{Prompt Adherence}, and \textit{Overall Quality}. Detailed implementation of the user study is provided in Appendix Sec.~\ref{sec:appendix_userstudy}.

\subsection{Qualitative Results}
\label{sec:result_quali}
Our in-context audiovisual model excels at preserving both visual and audio contexts, especially in challenging scenarios requiring precise temporal alignment and complex environmental sounds. As demonstrated in Fig.~\ref{fig:comp_duration}, our method    precisely aligns with the original video duration, avoiding visual rollback artifacts and duration mismatches observed in baseline methods.

The advantage of joint audiovisual modeling is further highlighted in scenarios involving visually related audio events. Fig.~\ref{fig:comp_nondialogue} and Fig.~\ref{fig:comp_event2} shows our method effectively leveraging visual context to distinguish segments needing translation from those to be preserved. As a result, it accurately maintains the timing of speech and essential non-dialogue audio elements such as pauses or laughter. In contrast, baseline methods, including the commercial tool HeyGen, entirely disregard these nuanced audiovisual cues.

Additionally, our method naturally adapts visual and audio content without relying on external face detectors. This strength is particularly evident under difficult conditions involving profile views, facial occlusions and non-human characters (Fig.~\ref{fig:comp_quality}). Baselines trained primarily on frontal-face reconstructions exhibit significant artifacts, such as faces incorrectly overlaying foreground objects, severely blurred lips, and failures resulting in unchanged copy-pasted input videos. Mask-based methods further fail entirely on stylized or non-human characters due to their reliance on explicit human-face priors.

Our unified audiovisual approach robustly generalizes to diverse and unconstrained scenarios, consistently maintaining accurate lip synchronization and stable visual quality, clearly outperforming existing baseline methods.

\subsection{Ablation Study}
\label{sec:ablation}

We analyze the necessity of our integrated design by jointly ablating training and lip augmentation on the synthesized dataset. We consider two variants: \textit{w/o Training}, a zero-shot in-context setting without learned adapters, and \textit{w/o Lip Augmentation}, where training data lacks phonetic diversity. As summarized in Table~\ref{tab:ablation}, both ablations expose the same failure mode: without learning prompt-driven audiovisual edits, the model defaults to reconstructing the source video. This reconstruction bias manifests as artificially high identity (ID-SIM) and voice similarity (V-SIM) but extremely poor prompt adherence, reflected by near-random WER. Removing lip augmentation further exacerbates this issue by reducing visual distinctiveness in training data, encouraging lip copy-paste behavior and degrading linguistic accuracy. In contrast, the full model achieves substantially lower WER and higher Lip Landmark Distance (LMD), confirming that it generates novel, prompt-aligned lip motion rather than replicating source trajectories. Together, these results demonstrate that training and lip augmentation are jointly necessary to break reconstruction bias and enable functional audiovisual dubbing.

\begin{table}[t]
\centering
\caption{\textbf{Ablation Study.} We evaluate identity preservation (ID-SIM), voice similarity (V-Sim), linguistic accuracy (WER), and lip distance (LMD).}
\label{tab:ablation}
\resizebox{\linewidth}{!}{
\begin{tabular}{@{}llcccc@{}}
\toprule
\textbf{Benchmark} & \textbf{Variant} 
& \textbf{ID-Sim $\uparrow$} 
& \textbf{V-Sim $\uparrow$} 
& \textbf{WER $\downarrow$} 
& \textbf{LMD $\uparrow$} \\ 
\midrule
\multirow{3}{*}{Easy} 
 & Full Model     
 & 0.8471 
 & 0.4235 
 & \textbf{0.3133} 
 & \textbf{0.0172} \\
 & w/o LoRA       
 & \textbf{0.9446} 
 & \textbf{0.8092} 
 & 0.9652 
 & 0.0086 \\
 & w/o Lip Aug    
 & \underline{0.9124} 
 & \underline{0.5621} 
 & \underline{0.5132} 
 & \underline{0.0094} \\ 
\midrule
\multirow{3}{*}{Hard} 
 & Full Model     
 & 0.6455 
 & 0.5578 
 & \textbf{0.2739} 
 & \textbf{0.0488} \\
 & w/o LoRA       
 & \textbf{0.7612} 
 & \textbf{0.8404} 
 & 0.9809 
 & 0.0432 \\
 & w/o Lip Aug    
 & \underline{0.7245} 
 & \underline{0.6832} 
 & \underline{0.5214} 
 & \underline{0.0469} \\ 
\bottomrule
\end{tabular}
}
\end{table}

\section{Conclusions}

\begin{figure*}[t]
    \centering    
    \includegraphics[width=0.9\linewidth]{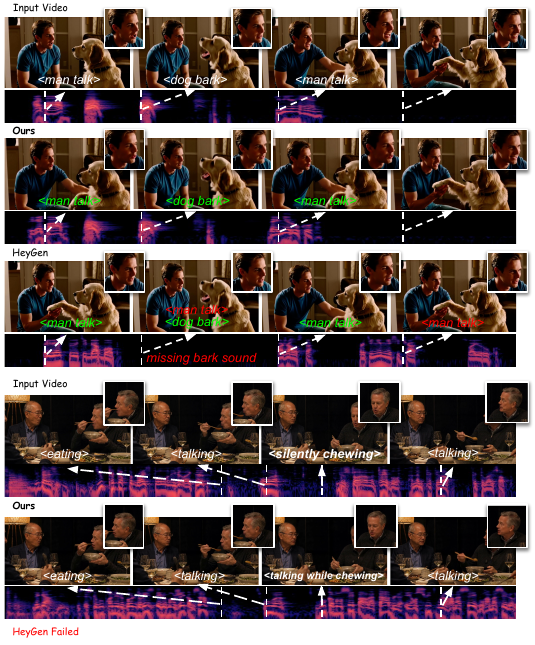}
    \vspace{-6mm}
    \caption{\textbf{Preservation of Non-Dialogue Events and Scene Grounding.} Our joint generative framework enables holistic scene modeling where visual dynamics and acoustic events co-evolve. \textbf{Top}: In the "dog barking" scenario, our method synchronizes the timing of environmental sounds with physical gestures, whereas baselines like HeyGen often omit or misalign these cues. \textbf{Bottom}: In the "eating while talking" example, our unified model elegantly manages cross-lingual duration mismatches. When translating the source English dialogue (9 syllables) into a longer French equivalent (11 syllables), the framework adaptively inserts speech frames during the speaker's chewing phase rather than when the mouth is obstructed by food. As demonstrated in the supplemental materials, the resulting output faithfully reproduces the acoustic texture of "talking through chewing," showcasing a bidirectional interaction between the scene's physical actions and the generated audio.}
    \label{fig:comp_event2}
\end{figure*}

We have presented a new way to approach video dubbing as a constrained audio–visual generation task. By operating directly within a joint audio–visual diffusion model, we show that translated speech, facial motion, and scene dynamics can be regenerated together under a unified generative prior. This formulation allows temporal structure, identity cues, and non-speech events to co-evolve naturally, resulting in dubbed videos that preserve both fine-grained lip synchronization and the broader semantic timing of the scene. Beyond robustness in challenging real-world settings, this work suggests that leveraging audio–visual foundation models offers a promising direction for moving dubbing beyond modular pipelines toward more holistic, context-aware generation.

Framing video dubbing as joint audio–visual generation introduces several technical challenges, including the lack of paired multilingual audiovisual data, the tension between preserving speaker identity and producing correct pronunciation, and the risk of information leakage when regenerating localized facial motion. In this work, we address these challenges while maintaining precise temporal alignment across speech, facial dynamics, and non-speech events within a single generative process.
\paragraph{Limitations} While the approach substantially improves overall dubbing quality, it does not yet perfectly preserve speaker voice identity in all cases, highlighting the need for stronger disentanglement between linguistic content and vocal style or more explicit identity supervision. 
\paragraph{Future Work} More broadly, this work underscores the potential of leveraging strong joint audio-visual generative priors for complex multimodal editing tasks and motivates future extensions to longer temporal contexts and richer conversational settings.

\begin{acks}
We thank Harel Cain, Mohammad Salama and Shachar Honig for valuable discussions, we thank Yoav Shtibelman and Rotem Banet for helping us create the video demo, and we thank Michael Kupchick, Noa Kotler, Andrew Kvochko, Amit Pintz and Alexey Kravtsov for the inference code support.
\end{acks}

\clearpage
\begin{figure*}[t]
    \centering

    \includegraphics[width=0.99\linewidth]{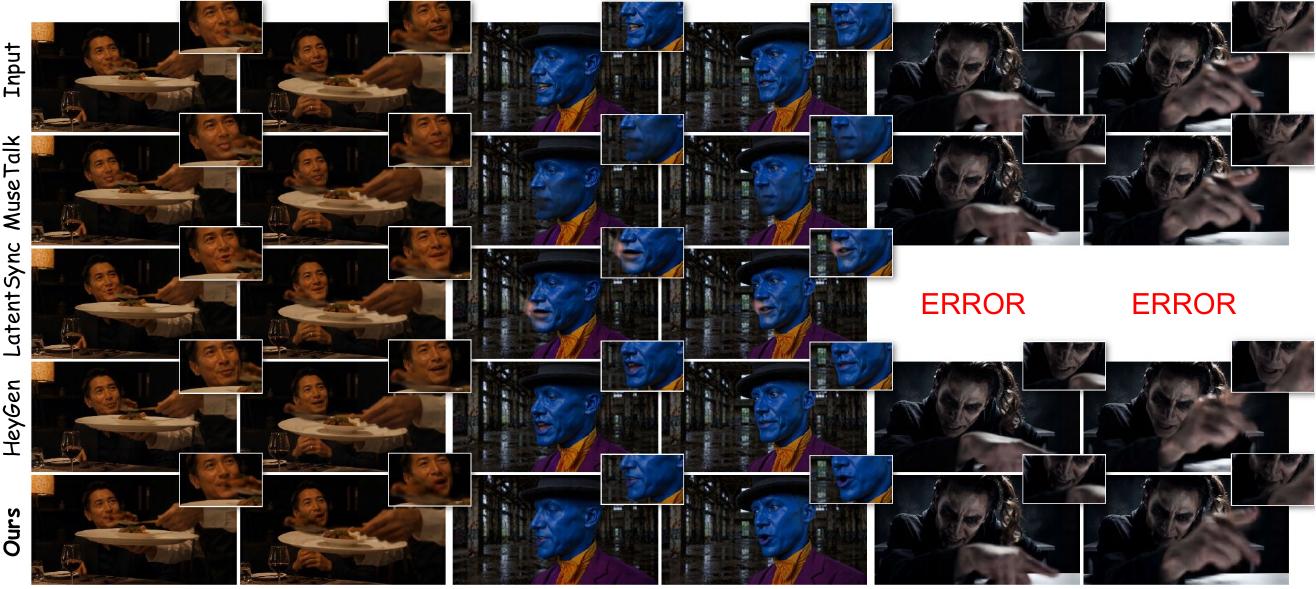}
    \vspace{4mm}

    \includegraphics[width=0.98\linewidth]{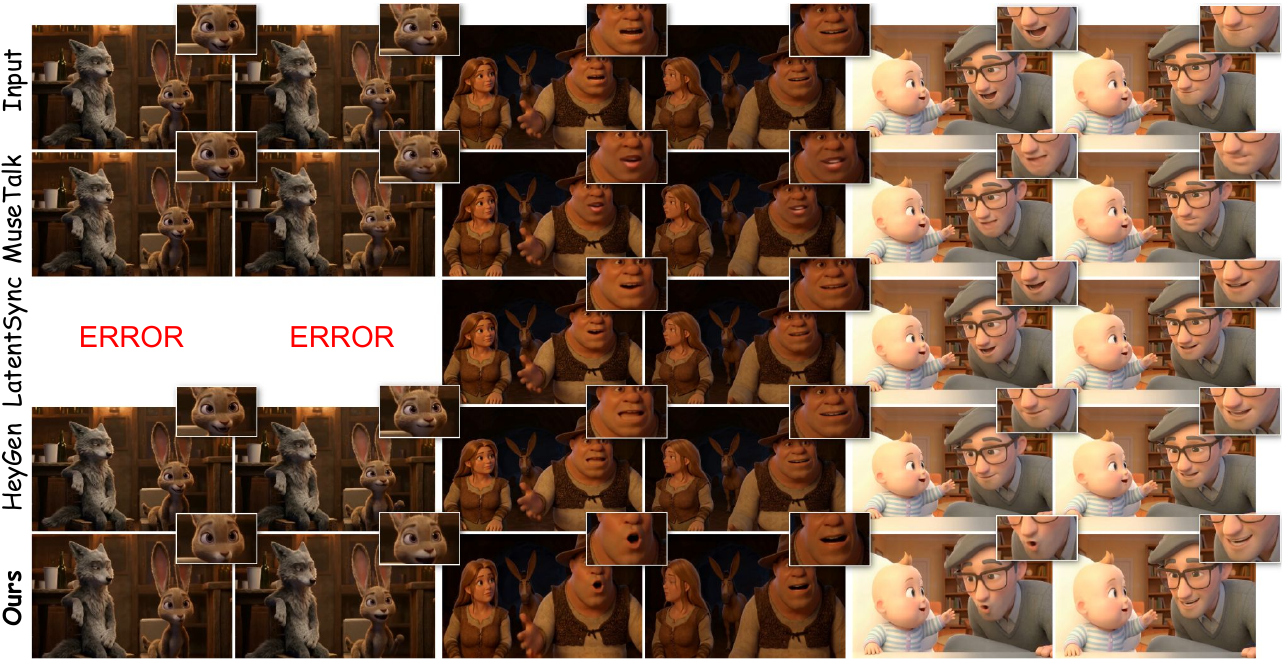}

    \caption{\textbf{Qualitative Comparisons.}
    \textbf{Top}: Profile views and occlusions.
    \textbf{Bottom}: Non-human scenarios.
    Baseline methods exhibit noticeable artifacts and often fail under these challenging conditions, while our method robustly preserves identity and visual coherence while synchronizing the lips with our generated translation audio.}
    \label{fig:comp_quality}
\end{figure*}
\clearpage

\bibliographystyle{ACM-Reference-Format}
\bibliography{main}

\clearpage
\begin{center}
  {\LARGE\bfseries Supplementary Material\par}
  \vspace{1em}
\end{center}

\pagestyle{plain}

\appendix

\section{Implementation Details}

\subsection{Training Data}
\label{sec:appendix_data}
We use Gemini to generate 100 structured multilingual prompts spanning seven languages: English, Spanish, Russian, French, German, Italian, and Greek. Each prompt specifies detailed visual attributes, temporal segments, and explicit language switches within a single video. An example prompt is provided below.

\begin{quote}
\small
\textit{
Hyper-realistic 8k, sharp focus, neutral lighting. Shot in a rainy alleyway at midnight with atmospheric cinematic lighting. Eleanor, a detective whose appearance matches the setting.
At [00:00--00:04]: Eleanor looks directly at the camera, speaking with urgency. She speaks in French, saying: ``La situation est devenue bien trop dangereuse pour que nous puissions rester ici.''
At [00:04--00:06]: SILENT PAUSE. She takes a deep breath.
At [00:06--00:10]: She lights a cigarette with shaking hands, shielding the flame. She speaks in English, saying: ``I explicitly told you never to come to this part of the city after midnight.''
}
\end{quote}

Given a generated language-switching video, we process the two halves independently. We detect the lip region using ArcFace and encode the masked video into latent space to obtain $F_{\text{mask}}$. In parallel, an all-black video is encoded to produce $F_{\text{empty}}$. The effective latent mask is computed as $|F_{\text{mask}} - F_{\text{empty}}| > \tau$, where $\tau = 0.1$. This mask is used to selectively apply noise to the lip region of the visual half, while the audio stream is fully noised. The noised audio-visual inputs are then jointly denoised using the same audio-visual diffusion model with a translated prompt. Notably, the lips mask is used only in the dataset generation stage. Later on in the LoRA training and inference, no mask is used.

For lip augmentation, we re-inpaint the same video using a randomly generated nonsensical speech prompt composed of letters (e.g., ``she is speaking `A$\ldots$B$\ldots$C' ''). The audio from the first inpainting pass and the video from the second pass are merged to form a context video. Although the audio and video are not temporally synchronized, the attention isolation mechanism limits cross-modal interference and mitigates adverse effects.

In data filtering process, besides semantic alignment using Qwen2-VL, linguistic correctness using Whisper and ArcFace for identity preservation. we use Lip Landmark Distance (LMD) and the Mouth Aspect Ratio (MAR) to quantify lip movement. The Lip Landmark Distance (LMD) measures structural deviation between reference and generated lip landmarks:
\begin{equation}
LMD = \frac{1}{T \cdot |L|} \sum_{t=1}^{T} \sum_{i \in L} \left\| \mathbf{P}_{i,t}^{ref} - \mathbf{P}_{i,t}^{gen} \right\|_2,
\end{equation}
where $T$ is the number of frames and $L$ denotes the set of 20 lip landmarks (indices 52--71).

The Mouth Aspect Ratio (MAR) at frame $t$ is defined as:
\begin{equation}
MAR_t = \frac{\left\| \mathbf{P}_{55,t} - \mathbf{P}_{61,t} \right\|_2}{\left\| \mathbf{P}_{52,t} - \mathbf{P}_{58,t} \right\|_2}.
\end{equation}

The MAR diversity ($MAR_{\text{div}}$) is computed as the standard deviation of MAR over the sequence:
\begin{equation}
MAR_{\text{div}} = \sqrt{\frac{1}{T} \sum_{t=1}^{T} \left(MAR_t - \overline{MAR}\right)^2}.
\end{equation}

Finally, the Quality--Diversity (QD) score is defined as:
\begin{equation}
QD = LMD \times MAR_{\text{div}}.
\end{equation}

\subsection{Benchmark}
\label{sec:appendix_benchmark}

We evaluate our framework using a comprehensive evaluation framework for video-to-video dubbing across two distinct tiers:

Standard Benchmark: This tier consists of 150 meticulously resampled pairs: 100 multilingual pairs from TalkVid  and 50 English-to-English pairs from HDTF. These datasets provide high-fidelity facial animations and natural speaking styles across English, German, Spanish, French, and Russian.

Challenging Benchmark: This tier comprises 50 high-complexity samples to stress-test robustness. It includes 25 edited movie clips sourced from YouTube featuring intense speech patterns (The Godfather, Pulp Fiction, The Mask, The Dark Knight) and 25 ``in-the-wild'' synthetic scenes. These samples involve low light conditions, profile views, occlusions, and non-human stylized characters.

\subsection{Training}
\label{sec:appendix_training}

We train a masked audio-video IC-LoRA adapter on the LTX-2 model using rank-128 LoRA modules applied to attention (\texttt{to\_k}, \texttt{to\_q}, \texttt{to\_v}, \texttt{to\_out.0}) and feed-forward (\texttt{net.0.proj}, \texttt{net.2}) layers across both video and audio branches. The training employs a dual learning rate strategy with $2 \times 10^{-4}$ for video modules and $1 \times 10^{-5}$ for audio modules to prevent overfitting in the audio domain. We implement masked loss training with a 10:1 ratio between foreground (\texttt{mask\_loss\_weight}$=1.0$) and background (\texttt{mask\_loss\_weight}$=0.1$) regions to focus learning on speaker-specific features while preserving scene context. The model is trained for 2,000 steps with batch size 1, gradient checkpointing enabled, and mixed-precision (bfloat16) on preprocessed video-audio pairs at $960 \times 544 \times 121$ resolution (25 FPS). Cross-attention masking between target video and reference audio is enabled to improve audio-visual alignment. We shifted logit-normal timestep sampling, and AdamW optimizer with linear scheduling and gradient clipping (\texttt{max\_norm}$=1.0$).

\subsection{User Study}
\label{sec:appendix_userstudy}
To evaluate our method, we conducted a comprehensive user preference study comparing our model against state-of-the-art baselines. We utilized a test set of 18 video samples, consisting of 6 easy samples (single speaker with static backgrounds) and 12 hard samples (cinematic scenes featuring complex motion, multiple speakers, and varied lighting). A total of 25 participants were asked to evaluate each sample across three dimensions: (1) \textbf{Lip Synchronization}, assessing how well the dubbed audio matches the facial motion; (2) \textbf{Prompt Adherence}, evaluating the correctness of the spoken words and pronunciation relative to the target dialogue; and (3) \textbf{Overall Preference}, considering factors such as acting quality, immersion, and professionalism. This resulted in a total of 450 pairwise comparisons and 1,350 individual data points. 

\subsection{Evaluation Metrics Implementation}
\label{sec:appendix_metric}
We report generation success rate (Succ) as the percentage of inputs for which a method produces an output video. In our evaluation, our method achieves 100\% Succ, while failures for competing methods occur when they cannot detect a face in the input. Identity preservation (CSIM) is computed as the cosine similarity between face-identity embeddings extracted from the generated video and the reference identity. Visual fidelity (FID) is the Fréchet distance between feature distributions of real and generated frames. Temporal coherence (FVD) is the analogous Fréchet distance computed in a video feature space to capture spatiotemporal consistency. Mouth Aspect Ratio diversity (MAR Div.) is computed as the temporal standard deviation of the Mouth Aspect Ratio (MAR), a normalized mouth-opening measure analogous to the Eye Aspect Ratio (EAR)~\cite{cech2016real}. Audiovisual synchronization (ASync) is computed using SyncNet, which measures a global synchronization offset for a video by averaging pairwise audio–visual embedding similarity across time.

\section{Additional Visualizations}

\subsection{Latent-Aware Fine Masking}
\label{sec:appendix_masking_example}

To demonstrate the existence of latent information leakage in training-free inpainting, we conduct an input-level corruption experiment by explicitly layering a green mask over the lip region of the test video before encoding. The corrupted video is then encoded into the Video VAE latent space, and diffusion is performed using a coarse latent mask corresponding to the lip area. As shown in Fig.~\ref{fig:latent_masking}, without Latent-Aware Fine Masking, the model produces pronounced green light artifacts around the mouth and lower face. These artifacts indicate that the green-masked lip information propagates into neighboring latent tokens during VAE encoding due to the large spatiotemporal receptive field, and is subsequently amplified during diffusion. In contrast, applying Latent-Aware Fine Masking removes the full latent region affected by this information spread, eliminating the artifacts and forcing the model to regenerate coherent, audio-aligned facial motion. This result confirms that naïve masking at either the input or latent level is insufficient, and that explicitly accounting for latent information propagation is critical for stable inpainting.

\begin{figure}[t]
  \centering
  \includegraphics[width=0.8\columnwidth]{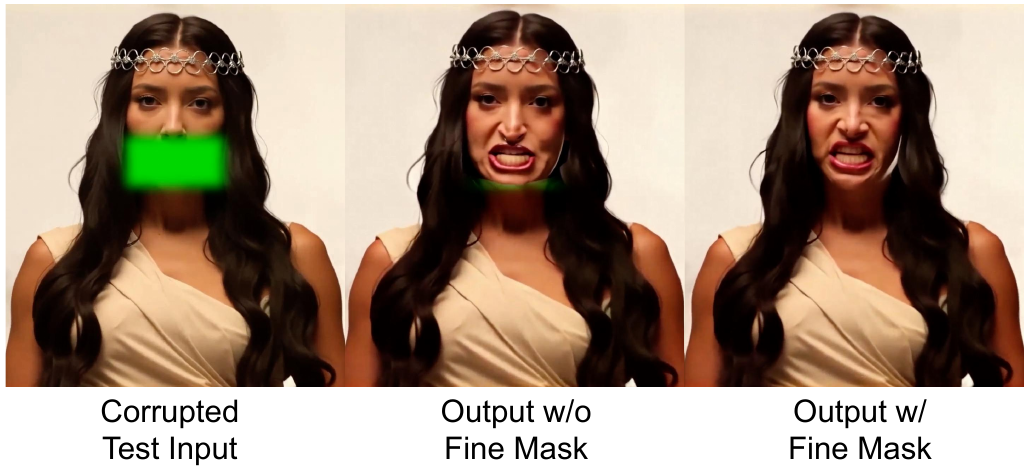}
  \caption{We overlay a green mask on the lip region of the input video and encode the corrupted video using the Video VAE. Left: Corrupted input. Middle: Output without Latent-Aware Fine Masking, showing pronounced green light artifacts around the mouth and lower face due to latent information leakage from the masked region. Right: Output with Latent-Aware Fine Masking, where leakage-induced artifacts are eliminated and coherent, audio-aligned facial motion is regenerated.}
  \label{fig:latent_masking}
\end{figure}

\subsection{Lip Augmentation}
\label{sec:appendix_augmentation_example}
The results in Fig.~\ref{fig:lip_augmentation} demonstrate that incorporating lip augmentation leads to substantially richer variation in reconstructed mouth movements, improving expressiveness and temporal realism. In contrast, inpainting without lip augmentation tends to collapse to similar lip shapes across frames, reducing motion diversity.

\begin{figure}[t]
    \centering
    \includegraphics[width=\linewidth]{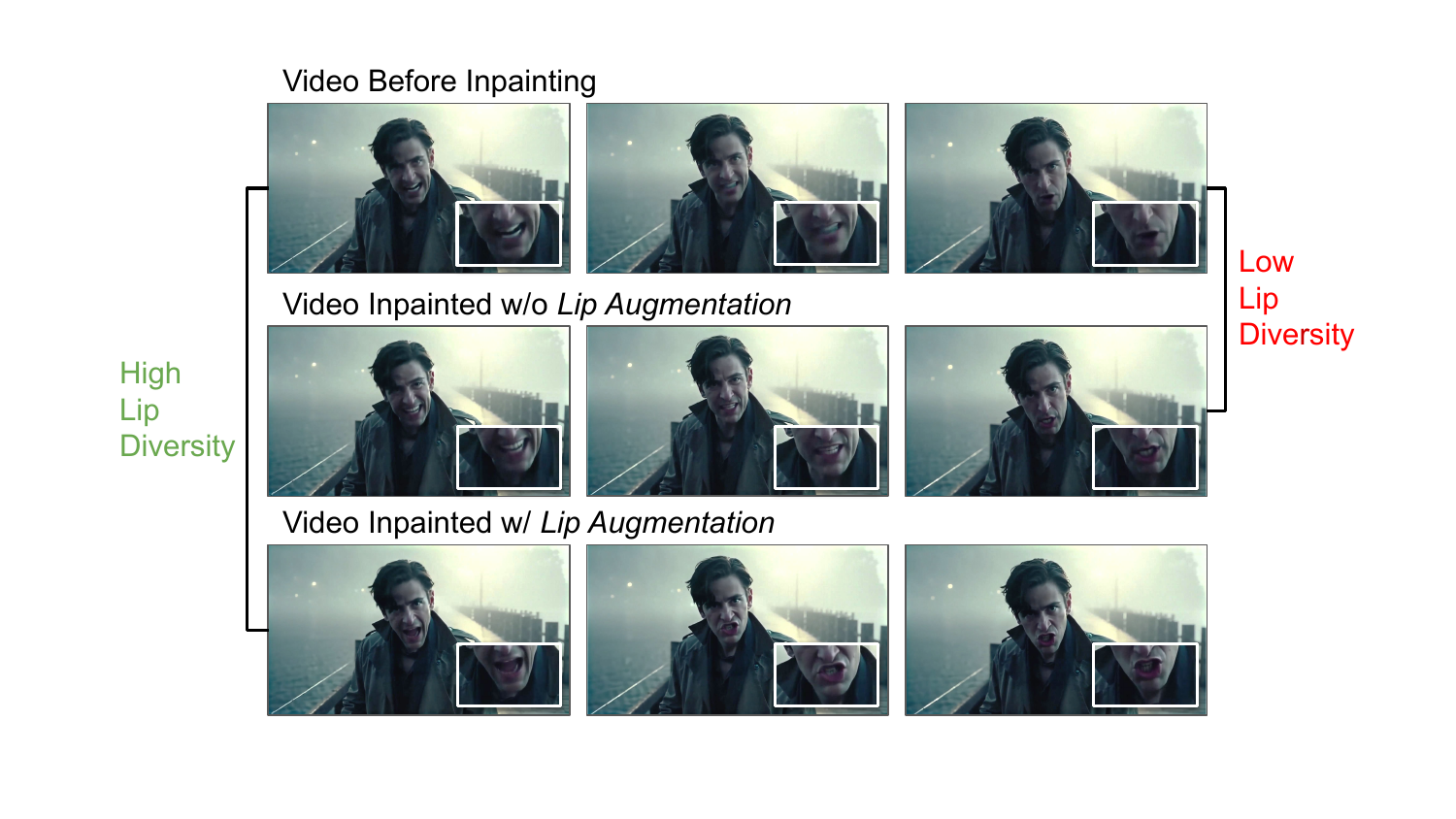}
    \caption{\textbf{Effect of Lip Augmentation on Data Generation Diversity}. We present visualizations on original video frames (top row), inpainted results without lip augmentation (middle row), and inpainted results with lip augmentation (bottom row). }
    \label{fig:lip_augmentation}
\end{figure}

\begin{figure}[t]
  \centering
  \includegraphics[width=\columnwidth]{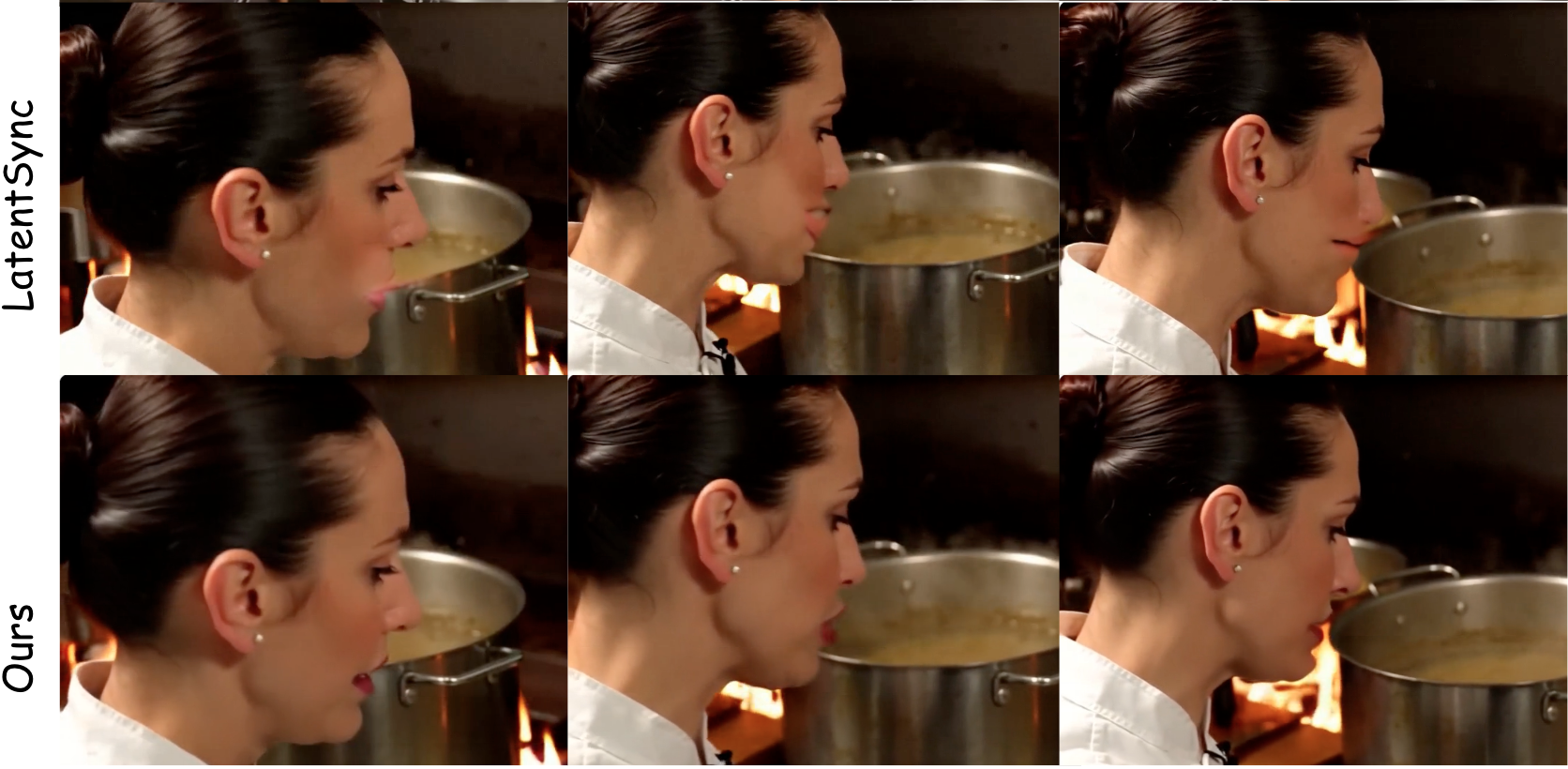}
  \caption{ASync metrics overfit to frontal videos. }
  \label{fig:async_overfit}
\end{figure}

\subsection{ASync metric problem}
\label{sec:appendix_async_example}
In Fig.~\ref{fig:async_overfit} unconstrained scenarios involving profile views, traditional metrics like SyncNet can be highly misleading. Masked-based baselines like LatentSync often reconstruct profile lips with a "frontal-facing" bias to minimize temporal offset, resulting in a misleadingly superior score of 0.00 despite visible artifacts. Our method achieves an ASync score (3.00) identical to the ground truth, demonstrating that our video results are physically consistent rather than metric-driven distortion.

\end{document}